\documentclass[preprint,amsmath,amssymb]{revtex4-1}
\usepackage{graphicx}
\usepackage{epsfig}
\usepackage{bm}
\begin{document}

\title{Unphysical artifacts are responsible for molecular orientation with half-cycle pulses}

\author{Juan Ortigoso}

\affiliation{Instituto de Estructura de la Materia, CSIC, Serrano 121, 28006 
Madrid, Spain}

\date{\today}

\begin{abstract}
Sub-cycle electromagnetic pulses in the Terahertz region are considered to be one of the best ways to orient molecules. The simplest option corresponds to half-cycles pulses in the sudden regime, for which the pulse duration is much shorter than the rotational period. However, no experimental demonstration exists so far. We argue that this kind of molecular orientation is an unphysical artifact that arises from a constant component of the electric field that is not a solution of Maxwell's equations, and therefore cannot propagate with the pulse. We present calculations based on Floquet theory that prove this claim.
\end{abstract}

\maketitle

\section{Introduction}\label{uno}

Rapid experimental developments of cooling methods and intense coherent sources has boosted an extraordinary interest in the understanding of phenomena related to the quantum control of molecular degrees of freedom \cite{cooling, cooling2}. In the last decade molecular alignment was one of the areas where impressive success was achieved, with the development of theoretical strategies \cite{ortigoso}, and the experimental demostration of several techniques \cite{seide1, seide1b}. However, molecular orientation has not progressed at the same speed.
Control of molecular orientation with half-cycle pulses (HCPs) has been intensively studied in the last few years promising that high orientation can be obtained in a very efficient way \cite{orien1, orien1b, orien1c}. Nonetheless, an experimental demonstration of the phenomenon has not been achieved yet. 

HCPs are time-dependent short  electromagnetic pulses in the THz region \cite{hcp}, that, ideally, have a single field maximum. The strong intensity of these pulses makes them appropriate to control diverse quantum processes in atoms and molecules \cite{ persson}. However, the time integral of the electric field must be zero for the pulse to be a solution of Maxwell's equations \cite{maxwell}. Thus, in practice HCPs are not trully unipolar. Nonetheless, these pulses can be represented by a strong unipolar part followed by a weak and very long tail where the field points in the opposite direction. This is required for the pulse to be propagable without much distortion. 

In theory, the unipolar part of a HCP interacts with the permanent dipole of molecules imparting them a sudden kick that transfers angular momentum breaking rotational parity. A molecule  described initially by a rotational eigenstate is converted by the pulse in a wave packet that contains rotational eigenstates of different parity, and for which the angle formed by the molecular axis and the field polarization vector is well defined. This view is based on a picture in which the HCP exerts a force on the  molecule only during the strong unipolar part, while the weak long tail does not significantly alter the relative populations of the wave packet components. For sudden interactions the orientation achieved at the end of the unipolar part of the HCP is usually very modest, but during the long tail the phases of the components of the wave packet sinchronize producing at some later time a large orientation.

Here, we show that the molecular orientation found with these theoretical models is consequence of an unphysical artifact, which is the fundamental reason for the failure to realize an experimental demonstration of the phenomenon, although arguments based on technical restrictions have been given too \cite{fcp2}. Thus, we prove that orientation is caused by the existence of a dc component in the half-cycle pulse. This component is not a solution of Maxwell's equations and in practice, it will be filtered out, along with other low frequency modes, during pulse propagation. 

Our calculations are based on Floquet theory, which allows us to remove in a simple way the effect of particular Fourier components of the pulse. We show that, by eliminating the dc contribution to the electric field, no orientation can be achieved in the sudden regime, although orientation can still be obtained for pulses of longer duration. In the next section we give a brief introduction to the Floquet formalism used in our calculations. In Section \ref{tres} we present our main results. In Sect. \ref{cuatro} we present our conclusions.

\section{Computational model}\label{dos}

\subsection{Floquet formalism} \label{dosa}

The time evolution of a physical system driven by a time-dependent periodic Hamiltonian can be obtained by employing the Floquet approach \cite{shirley}. This method is especially suited to study the interaction of a molecule with a train of electromagnetic pulses \cite{ortigoso2}. However, it can be used also to analyze systems driven by a single pulse, and that therefore are not periodic. This can be done by assuming that the pulse is a member of a periodic sequence, whose repetition period, $T$, is given, at least, by the effective duration of the solitary pulse.

The time evolved wave function is  $\Psi(t)=U(t,t_{0})\Psi(t_{0})$, where $t_{0}$ can be chosen as the time when the pulse is launched, and the time propagator is formally given by the expansion

\begin{equation}
U(t,t_{0})=\sum_{n} c_{n} e^{-i \lambda_{n}(t-t_{0})} \chi_{n}(t')|_{t=t'} \;,
\label{floquet1}
\end{equation}

\noindent
where $\lambda_{n}$ and $\chi_{n}$ are the eigenvalues and eigenvectors respectively of a Floquet Hamiltonian, ${\cal F}(t')=H(t')-i\partial/\partial t'$, that acts in an extended Hilbert space where time, $t'$, is treated like a spatial coordinate \cite{sambe}. The coefficients $c_{n}$ are given by the overlaps

\begin{equation}
c_{n}=\langle \chi_{n}(t'=t=t_{0})|\Psi(t_{0})\rangle\;.
\label{floquet2}
\end{equation}

In practice, instead of diagonalizing the Floquet Hamiltonian to obtain Eq. (\ref{floquet1}) it is more efficient, from a computational point of view, dividing  $(t, t_{0})$ into smaller time intervals of $\tau$ duration. Doing so, the propagator $U(t,t_{0})$ can be calculated as the product \cite{peskin},

\begin{equation}
\prod_{m=1}^{M} U\left[m \tau,(m-1)\tau \right] \;,
\label{pccp1}
\end{equation}

\noindent
where \cite{shirley, peskin}

\begin{equation}
U\left( m\tau,(m-1)\tau\right)=\sum_{n=-\infty}^{\infty} e^{2\pi i n m \tau/T}
\left\{I\delta_{n,0}-i\frac{\tau}{\hbar}\left[F(x)\right]_{n,0}
-\frac{1}{2} \frac{\tau^{2}}{\hbar^{2}}\left[F^{2}(x)\right]_{n,0}+\cdots \right\} \;.
\label{moiseyeveq}
\end{equation}

\noindent
The matrix $F$ depends on the spatial variables $x$. For small enough $\tau$ the Fourier expansion in $n$ can be truncated to a few elements. Apart from its efficiency, this method gives direct access to the effect of the different Fourier components of the  potential, since the time-dependent part of the Floquet Hamiltonian can be factorized as $V(x, t')=V(x) V(t')$,

\begin{equation}
[F_{ij}(x)]_{n0}=V_{ij}(x)\frac{1}{T}\int_{t_{0}}^{t_{0}+T} e^{-2 \pi i n t'/T} V_{ij}(t') dt' \;,
\label{equnueva}
\end{equation}

\noindent
which gives the $ij$ element of the matrix $F$, where $i$, $j$ are spatial basis functions.

\subsection{Rotating molecule interacting with a half-cycle pulse}\label{dosb}

Previous theoretical models are based on HCPs consisting of an unipolar part followed by an infinite tail of infinitesimally small negative intensity such that $\int_{-\infty}^{\infty} E(t)dt=0$.  Since the field at the tail is so weak it does not interact with the molecules, which can be considered to be under field-free conditions. Thus, any possible orientation effect comes from the symmetric unipolar part. 

For a linearly polarized half-cycle periodic pulse train, the unipolar part can be written, for example,
as $E(t)=E_{0}\exp(-(t-t_{0})^{2}/\sigma^{2})\cos(\omega t)$, where $\sigma$  is related to the Gaussian half-width, and $\omega$ is in the THz region. 
The Floquet Hamiltonian (in reduced units) for a rotating linear molecule in the presence of such an HCP can be written \cite{ortigoso2} as

\begin{equation}
{\cal F}(\theta, t')={\bf{\rm  J}}^{2}-\frac{\mu E_{0}}{B} \exp(-t'^{2}/\sigma^{2}) \cos (\omega t') \cos\theta - i\frac{\partial}{\partial t'}\;,
\label{floqueth}
\end{equation}

\noindent
where ${\bf{\rm J}}$ is the angular momentum vector, $B$ the rotational constant, $\mu$  the permanent dipole moment,  and $\theta$ is the angle between the polarization vector of the field and  the internuclear axis.  The  elements of the matrix representation of ${\cal F}$, that enter into Eq. (\ref{moiseyeveq}) are 

\begin{equation}
[F_{JJ'}(\theta)]_{n,0}=\langle JM|\cos\theta|J'M'\rangle \frac{1}{T}\int_{-T/2}^{T/2} \exp(-2 i \pi n t'/T)   \exp(-t'^{2}/\sigma^{2}) \cos (\omega t') dt' \;,
\label{matrixelemen}
\end{equation}

\noindent
where the rotational matrix elements are nonzero only for $J'=J\pm 1$, and $M'=M$ \cite{hougen}:

\[
\langle J-1|\cos \theta|J\rangle=\sqrt{\frac{(J+M)(J-M)}{(2J-1)(2J+1)}} \;,
\]

\begin{equation}
\langle J+1|\cos\theta|J\rangle=\sqrt{\frac{(J+M-1)(J+M+1)}{(2J+1)(2J+3)}} \;.
\label{matrixe1}
\end{equation}   

\section{Results}\label{tres}

\subsection{Orientation with a hypothetical model pulse}\label{tresa}

Traditionally, achievement of molecular orientation with time-dependent fields has been considered more difficult than the well understood molecular alignment because it requires breakdown of inversion symmetry \cite{bretis, bretisb, twocolor, twocolor2, twocolor2b}.  It is frequently argued that any orientational effect in molecules requires an external field asymmetric in time. Thus, the highly asymmetric temporal structure of HCPs seems especially appropriate to achieve orientation.

\begin{figure}
\centering
\epsfig{file=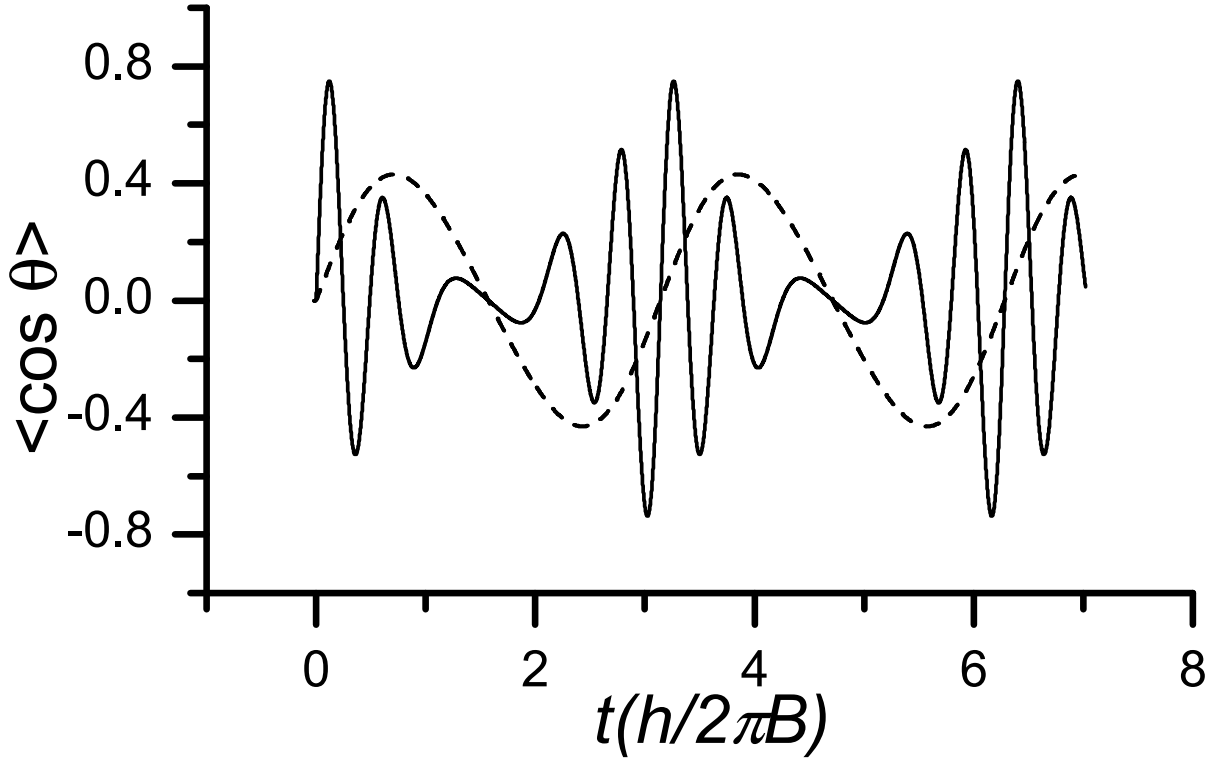,width=\columnwidth}
\caption{Instantaneous molecular orientation, $\langle \cos \theta \rangle$, as a function of time, in reduced units, for a molecule initially in the eigenstate $|J=0, M=0\rangle$. Results of applying  two different half-cycle pulses are shown. Common parameters for both cases are $\sigma=0.006 (\hbar/B)$, $T=0.04 (\hbar/B)$, $\omega=200 (B/\hbar)$. Solid line is for $\mu E_{0}/B=1000$, and dashed line corresponds to $\mu E_{0}/B=100$. The calculations were obtained using a Floquet operator, Eq. (\ref{moiseyeveq}), built with 15 rotational eigenstates. The time propagation after the end of the pulses was done under field-free conditions.}
\label{figure1}
\end{figure}

Available HCP pulses of enough strength are short compared to typical rotational periods.  Therefore, the dynamics is given by the sudden approximation. Under these circumstances, orientation is achieved only after the pulse is over.  Fig. \ref{figure1} shows the instantaneous orientation, as a function of time in reduced units (see figure caption), produced by two half-cycle pulses with different intensity. When the unipolar part of the pulses ends no orientation has been achieved. However, the interaction has already broken the parity symmetry and the rotational wave function contains eigenstates of different parity. For the pulse with lower intensity (dashed line), the wave function at the end of the unipolar part is essentially given by $\Psi(t=T/2)\approx 0.91|J=0, M=0\rangle+0.40i|J=1, M=0\rangle$. During the long tail the molecules orient progressively to reach a maximum. For example, at $t=1 \hbar/B$, i.e.,  near the maximum of the dashed line in Fig. \ref{figure1} the wave function is $\Psi(t=T/2)\approx 0.91|J=0, M=0\rangle-(0.17+0.36 i)|J=1, M=0\rangle$. After molecules reach maximum orientation the components of the wave function dephase again and orientation is lost. The time scale of this phenomenon depends, for a particular HCP, on the magnitude of the rotational constant. For example, a molecule with  $B=10$ cm$^{-1}$ loses its maximum orientation in less than 100 fs, while a molecule with $B=0.1$ cm$^{-1}$ remains well oriented arount 50 ps, which may be comparable to the tail duration. This imposes an additional restriction to the lenght of the HCP tail if successive pulses, forming a train, are used, as discussed in the next section.

\subsection{Effect of different pulse components on molecular orientation}\label{tresb}

The previous results are based on the assumption that the shape of the strong unipolar part of the pulse does not change during propagation. However, it is well known \cite{prekaplan} that the temporal profile of THz beams reshapes during propagation in free space. This indicates that the quantitative analysis of experiments involving the interaction of half-cycle pulses with molecules requires the precise determination of the actual shape of the pulse that effectively reachs the sample. 
Here, our goal is less technical. We are interested in a proof of principle that does not require to know the particular shape that a given pulse adopts after free-space propagation or after passing through optical elements. The only concept needed for our purposes is that the dc component of the pulse is not a solution of Maxwell's equations and consequently it is not propagable. Thus, the pulse reaching the sample does not contain such static component regardless of its temporal shape. In the same way, other low-frequency components are diffracted away during propagation.
Next, we present calculations that prove this claim.

As argued in the Introduction, Floquet theory offers additional advantages over other propagation methods in terms of understanding solutions to the time-dependent Schr\"odinger equation. The method gives direct access to the matrix elements of the different field components, and their influence can be readily assessed. 

\begin{figure}
\centering
\epsfig{file=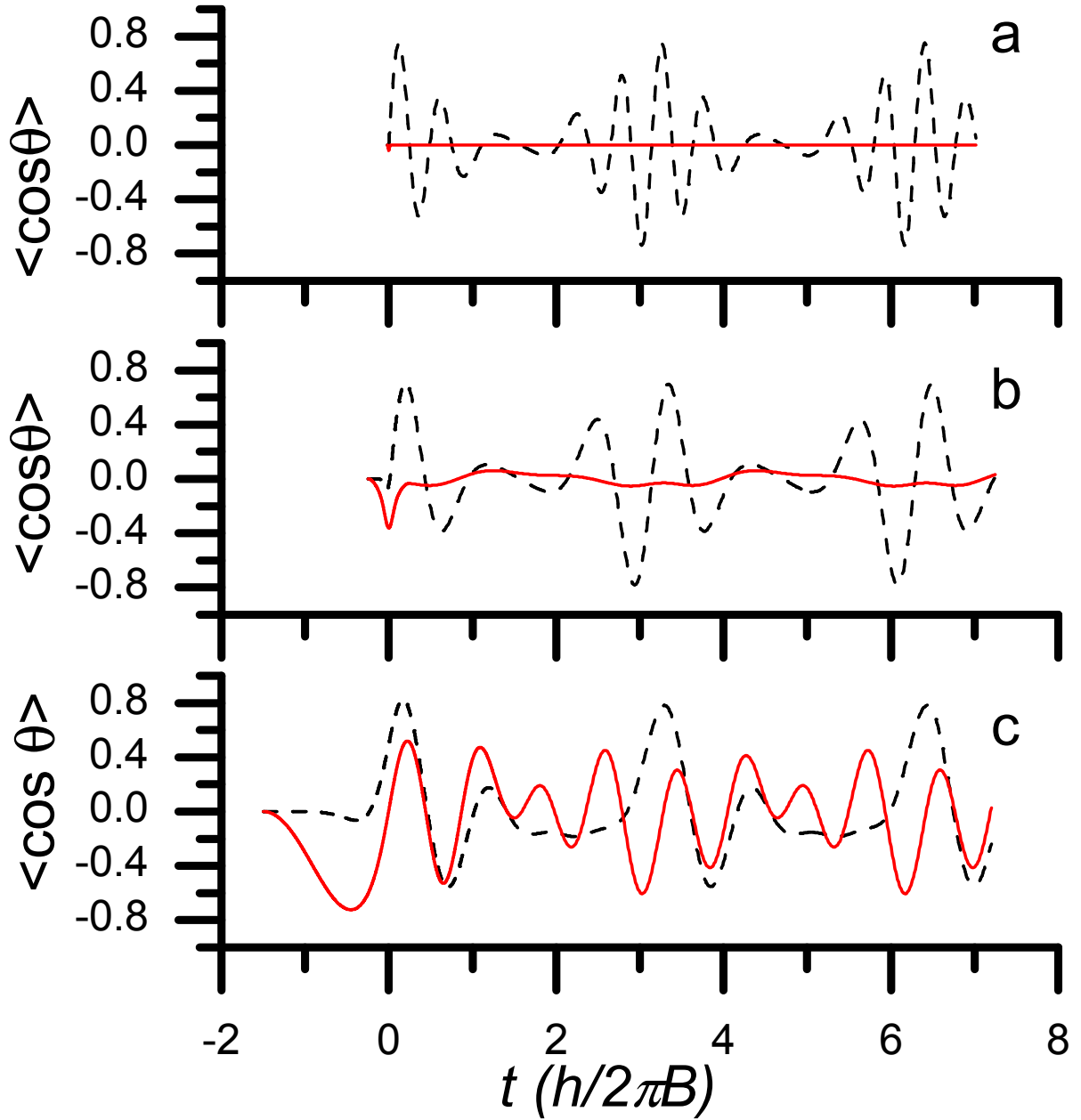,width=\columnwidth}
\caption{Orientation, $\langle \cos \theta \rangle$, with HCP pulses for three different regimes showing the effect of the static component of the electric field. Solid lines correspond to calculations in which the dc component of the field has been removed, while dashed lines have been calculated including it. The parameters, in reduced units, for thee three pulses are:  
(a) $\sigma=0.006$, $T=0.04 $, $\omega=200 $, $\mu E_{0}/B=1000$, (b) $\sigma=0.07$, $T=0.5$, $\omega=30$, 
$\mu E_{0}/B=100$, and (c) $\sigma=0.45$, 
$T=3$, $\omega=3$, $\mu E_{0}/B=12$. The initial time for the time propagation is $t_{0}=-T/2$, which varies with the pulse duration. The time propagation after the end of the pulses was calculated under field-free conditions.} 
\label{figure2}
\end{figure}

In effect, any dc field component can be eliminated by zeroing the elements with $n=0$ in Eq. (\ref{moiseyeveq}). The result of this procedure is shown in Fig. \ref{figure2}, for pulses corresponding to three different regimes according to the duration of the unipolar part of the HCP. For example, for a molecule with $B=1$ cm$^{-1}$, panel b corresponds to a pulse with $\sigma\approx 400$ fs. Dashed lines show the results of calculations in which the non-propagable static component of the electric field has been retained. In this case large orientations are obtained regardless of the pulse duration. On the other hand, solid lines show the results of calculations in which the effect of the static component has been eliminated. In this case, panel a shows that for the shortest pulse, corresponding to the sudden regime,  orientation is fully supressed. Panel b shows results for a pulse ten times longer, which is well beyond the sudden regime. In this case, a maximum in orientation is achieved at some time during the pulse. Although strongs peaks are visible when using the full HCP, orientation is basically suppressed after filtering out the dc component. A small 
amount of orientation remains, indicated by the negative peak just at the center of the unipolar part of the pulse, but it dissapears by the end of the pulse and does not reappear during the tail. This phenomenon is due to the particular form of the rotational wave packet excited by the pulse without dc component. While the complete HCP mixes up to seven rotational eigenstates by the end of the pulse, our results show that the wavefunction obtained with the filtered HCP is $\Psi(T/2) \approx (0.78-0.58i)|J=0, M=0\rangle+0.16|J=2,M=0\rangle$ with a very small contribution of the $J=1$ state that is not enough to give further orientation. However, the wave function at the time where a significant orientation is achieved ($t\approx 0$) contains an eight percent contribution of the odd state, enough for producing a $\langle \cos \theta\rangle (t=0)\approx -0.36$. 
Finally, results for pulses that last several ps are shown in panel c. In this case, the electric field has a dc component too, but its relative contribution is much less important than for the sudden regime. In this regime orientation is possible even if the dc component is filtered out.
 
\section{Conclusions} \label{cuatro}

Molecular orientation is due to the hybridation of rotational eigenstates of different parity by an external field. In spite of the broad-band width of half-cycle pulses, only low frequency components close to rotational spacings can significantly couple rotational states. However, real HCPs cannot have such a  dc component because it is not a solution of Maxwell's equations. Analytical calculations showing changes after propagation, for pulses with arbitrary temporal profiles were presented by Kaplan \cite{kaplan}. Also, pulse reshaping due to the effect of lenses and mirrors was studied by You and Bicksbaum \cite{distortion}. Pulse propagation in free space and through far-infrared optics elements are necessary to focus the radiation onto, for example, a molecular sample. Therefore, if the interaction of the pulse with the molecules takes place after having propagated a certain distance, it will be unable to produce any noticiable orientation, because far from the source, the dc and low frequency components are filtered out distorting the field. These field components diffract away as soon as they leave the emitter \cite{distortion}.  This notion inmediately suggests that short propagating half-cycle pulses will hardly have any effect on molecular orientation. 

Our results prove that the molecular orientation found in theoretical studies is caused precisely by the dc component of the unipolar part of the pulse.  Thus, although previous theoretical studies had attributed molecular orientation, in the sudden regime, to the asymmetric structure of HCPs, the actual mechanism at work in those models is the existence of an unphysical Stark effect. 

The previous discussion concerns excitations with a single pulse. Further limitations arise for schemes based on sequences of pulses that impart further kicks at each time the molecules start to lose their maximum orientation \cite{hcptrain, hcptrainb}. These strategies were devised to maintain persistent molecular orientation. However, as we showed in Sect. \ref{tresa} they only can work for heavy molecules for which the time to achieve the maximum orientation is comparable to the duration of the long HCP tail. Therefore, a new pulse can be sent at the correct time without interfering with the previous one. On the other hand, light molecules go through several orientation/misorientation cycles during the tail. Thus,  the orientation will be destroyed before the arrival of a new pulse.  This shows that the long-tail requirement is a serious obstacle to obtain persistent orientation with sequences of pulses except for heavy enough molecules.  

On the other hand, fruitful applications of half-cycle THz pulses have been demonstrated. For example, a protocol for steering Rydberg electrons towards selected final states was realized with a tailored sequence of HCPs \cite{transport}. Also, numerical simulations have demonstrated that control of dissociation of small molecules \cite{diso, iso} or isomerization \cite{isomer} are possible.
However, the mechanism at work in these cases is fundamentally different to that of rotational control here studied, due to the peculiar structure of rotational energy levels. The Floquet approach used in the present work shows that terms in the time propagator with $n=0$ are responsible for the orientation found in theoretical calculations. Elimination of these terms in the frequency spectrum and further reconstruction of the time envelope produce a pulse that is equivalent to a HCP for which the unipolar part in the positive direction is off-set by a negative weak field. This type of pulses has been successfully employed to demonstrate control of electron localization in molecules \cite{persson}.

The main conclusion of our work indicates that further studies should abandon the notion of HCP-induced molecular orientation in the sudden regime. Alternatives based on the use of longer pulses or proposals based on the use of few-cycle pulses \cite{chin, bandrauk, fcp, fcp2} are not affected by our results and practical implementations should be actively pursued.

\section*{Acknowledgments} 

Financial support from the Spanish Government through the MICINN project  FIS2010-18799, is acknowledged.

\end{document}